\documentclass[reprint,pre]{revtex4-1}
\pdfoutput=1
\usepackage{subfigure}
\usepackage{latexsym}
\usepackage{dsfont}
\usepackage{url}
\usepackage{times}
\usepackage{graphicx}
\usepackage{amsfonts}
\usepackage{amsmath}
\usepackage{ifthen} 
\usepackage{verbatim}
\usepackage{color}
\usepackage{hyperref}
\hypersetup{
    colorlinks=true,
    linkcolor=blue, 
    citecolor=blue,
    urlcolor=blue,
    pdftitle={Fast Generation of Spatially Embedded Random Networks},
    breaklinks
}

\def\equationautorefname~#1\null{%
  (#1)\null
}

\usepackage{epstopdf}

\usepackage{algcompatible,newfloat}
\DeclareFloatingEnvironment[
    fileext=loa,
    listname=List of Algorithms,
    name=ALGORITHM,
    placement=tbhp,
]{algorithm}

\newcommand{\vect}[1]{{\mathbf #1}}
\newcommand{\bx}{\vect{x}}
\newcommand{\norm}[2]{\left|\left| #1 \right|\right|_{#2}}
\newcommand{\R}{{\mathbb R}}

\newcommand{\titlestr}{Fast Generation of Spatially Embedded Random Networks}

\newcommand{\ie}{{\em i.e., }}
\newcommand{\eg}{{\em e.g., }}

\newcommand{\anonymize}[1]{anonymous}
\newcommand{\E}{\ensuremath{\mathbb{E}}}
\newcommand{\naive}{na\"{i}ve }

\newlength{\figurewidth}
\begin{document}
\title{\titlestr}
\author{Eric Parsonage and Matthew Roughan}
\affiliation{$<\!\!\!\{$eric.parsonage,matthew.roughan$\}$@adelaide.edu.au$>$ \\
  ARC Centre of Excellence for Mathematical \& Statistical  Frontiers, \\
  School of Mathematical Sciences \\ 
  University of Adelaide}
 
\begin{abstract}
  Spatially Embedded Random Networks such as the Waxman random graph have
  been used in a variety of settings for synthesizing networks.
  However, little thought has been put into fast generation of these
  networks. Existing techniques are $O(n^2)$ where $n$ is the
  number of nodes in the graph. In this paper we present an
  $O(n + e)$  algorithm, where $e$ is the number of edges.
\end{abstract}


\maketitle

\section{Introduction}

Random graphs are frequently used as the underlying model in fields
such as computer networking, biology and physics, but increasingly 
the datasets we wish to model involve  large numbers of elements.
This is problematic because common methodologies that investigate the 
properties of such large graphs involve the generation of random graphs 
as examples in order to investigate their asymptotic behavior's and 
though there are many works on the analysis of random graphs, 
there are relatively few on how to generate large graphs {\em efficiently}.

This paper is concerned with efficient generation of large Spatially
Embedded Random Networks (SERN). These arose soon after the
Gilbert-Erd\H{o}s-R\'enyi (GER) random graph
\cite{gilbert59:_random_graph,erdos60} with the {\em random plane
  network} proposed by
Gilbert~\cite{gilbert61:_random_plane_networ}. However, the most cited example is the Waxman random graph
\cite{b.m.waxman88:_graph}, with the original paper being cited several thousand times.

The GER random graph links every pair of vertices independently with a
fixed probability, whereas the Waxman graph reflects that in real
networks longer links are often more costly or difficult to construct,
and their existence is therefore less likely. It links nodes $i$ and $j$
with a probability given by a function of the distance $d_{i,j}$
between them. The form chosen by Waxman was the negative
exponential
\begin{equation}
  \label{eq:waxman_prob} 
  p(d_{i,j}) = q \, e^{- s d_{i,j}},
\end{equation}
for parameters $q,s \geq 0$.  A Waxman random graph is generated by
randomly choosing a set of points in a section of the plane (usually
the unit square), and then linking these points independently
according to their distance.

The idea of a distance-based probability of connection has been
generalized in SERNs \cite{PhysRevE.76.056115,kosmidis08:_struc}. In
these the metric space in which the points are embedded is
generalized, as is the distance function, but the underlying concept
is identical. There are many examples of SERNs, including random plane networks, random geometric graphs, spatial networks, range-dependent random graphs, random connection mod- els, random distance graphs, and partially structured random graphs.

Generation of synthetic random graphs is one of the basic requirements
for modeling. Synthetic graphs allow consideration
questions such as ``how will a network behave if it grows?'' The
model overlaying the graph may be quite complex, such as a
routing protocol in computer networking. This means analysis of the
graph may not provide much insight whereas the ability to simulate the
model may.  Ability to generate a suite of such graphs which match
some characteristics of real graphs allows us to test not just
predictions, but also their sensitivity to the underlying assumptions
(for instance the choice of parameter values).  The facility to
synthesize graphs is also needed in estimation procedures such as
Approximate Bayesian Computation
(ABC)~\cite{csilery10:_approx_bayes_comput_abc}.

At present, the only algorithms available for generating SERNs use
the approach of generating the node locations, calculating the
distances, and then generating edges using a series of Bernoulli
trials, which is $O(n^2)$ in the number of nodes. We call this the {\em
  \naive} algorithm.

However, many realistic graphs are very sparse in the sense that $O(e)
< O(n^2)$,  the number of edges grows more slowly that the
number of potential edges. So for many real-world examples an $O(n^2)$
algorithm is highly inefficient.

Here we develop a fast, efficient method for creating large sparse
SERNs. Our method takes $O(n + e)$ computation, which is the best
possible computation time for an exact method. We have used the method
to generate graphs with up to a billion nodes.

We also demonstrate a multi-threaded implementation that shows that
the method parallelizes. 

Code implementing the algorithm is available at
\url{github.com/lamestllama/conSERN}. 



\section{Background} 

SERNs \cite{PhysRevE.76.056115,kosmidis08:_struc} constitute a large
class of useful random-graph models, including Gilbert's random plane
network~\cite{gilbert61:_random_plane_networ} (also known by other
names such as the {\em random geometric graph}
\cite{penrose03:_random_geomet_graph}) and the Waxman random graph
\cite{b.m.waxman88:_graph}.  The Waxman graph has been used in many
settings from computer networks to biological cell networks, typically
to synthesize random networks.
We demonstrate our approach with this particular SERN but the reader
should keep in mind that our implementation already caters for the
general case.

We are not aware of any general tools to generate wide-classes of
SERNs, but there are a number that have been designed for
generating Waxman random
graphs~\cite{Medina:2001:BAU:882459.882563,zegura97:_quant_compar_of_graph_based,Magoni:2002:NSN:882460.882599,magoni01:_influen_networ_topol_protoc_simul,weisser10}. 
None seriously consider how to generate these graphs quickly.
\begin{itemize}

\item NetworkX \cite{networkx}, 
  aSHIIP~\cite{weisser10},
  NEM~\cite{Magoni:2002:NSN:882460.882599,magoni01:_influen_networ_topol_protoc_simul}
  and GT-ITM~\cite{zegura97:_quant_compar_of_graph_based} all generate
  the graph using the \naive $O(n^2)$ algorithm.

\item The Matlab Waxman graph generator \cite{kaj05:_stoch_matlab}
  also executes a vectorized \naive algorithm.

\item BRITE~\cite{Medina:2001:BAU:882459.882563} has two algorithms,
  but both appear to make serious deviations from the standard Waxman
  model in order to generate connected graphs. Also, although one
  approach is technically $O(e)$ it uses an rejection sampling
  approach that can take hours to generate even small networks.
  

\end{itemize}
All of the methods that generate true Waxman graphs are $O(n^2)$ in
computation time \cite{lothian13:_synth_graph}, and the vectorized
Matlab algorithm is $O(n^2)$ in memory as well.

In modern problems, networks of millions of nodes are common,
and billion node networks exist. For instance, FaceBook claims (as of
July 2015) over a billion active users, who form part of a large
graph. As network modeling moves towards encompassing such graphs, the
need for synthesizing very large graphs increases.

There are also some subsequent algorithms that require us to generate
a large number of random graphs. For instance
ABC~\cite{csilery10:_approx_bayes_comput_abc} requires a large number
of synthesized graphs, over a wide range of parameters. The
requirement means we need an efficient generator.

\subsection{Mathematical formalities}
\label{sec:formalities}

A graph (or network) consists of a set of $n$ vertices (we shall
synonymously refer to them as nodes), which, without loss of
generality, we label ${\cal V}=\{1,2,\ldots, n\}$, and edges (or
links) ${\cal E} \subset {\cal V} \times {\cal V}$. We are primarily
concerned here with undirected graphs (though much of the work on random
graphs is easy to generalize to directed graphs).  We say that two
nodes $i$ and $j$ are {\em adjacent} or {\em neighbors} if $(i,j) \in
{\cal E}$.

 
The GER random graph~\cite{gilbert59:_random_graph,erdos60}, $G_{n,p}$
of $n$ vertices is constructed by assigning each edge $(i,j)$ to be in
${\cal E}$ independently, with fixed probability $p$.  A SERN
generalizes this by making the probability of each edge dependent on
the distance between the two nodes.

Formally, we create a SERN by placing $n$ nodes randomly within some
defined region $R$ of a metric space $\Omega$ with distance metric
$d(x,y)$. Each pair of nodes is made adjacent independently, with link
probability given by a function of distance $d_{i,j} = d(x_i, x_j)$
between nodes $i$ and $j$. For instance we could define a space,
with one of the standard distance metrics
\begin{itemize}
\item Euclidean: $d_{i,j} = \norm{\bx_i - \bx_j}{2}$,

\item Manhattan: $d_{i,j} = \norm{\bx_i - \bx_j}{1}$,

\item Discrete: $d_{i,j} =  \norm{\bx_i - \bx_j}{0}$, 

\item Max: $d_{i,j} = \norm{\bx_i - \bx_j}{\infty}$, 

\end{itemize} 
and one of the following link probability functions:
\begin{itemize}

\item Waxman: $p_{i,j} = q\, e^{-sd_{i,j}}$, where $s \in [0,\infty), q
  \in (0,1]$, \cite{b.m.waxman88:_graph}\footnote{
Note that the parameterization \autoref{eq:waxman_prob} differs from
much of the literature on Waxman graphs. We chose to do this as
unfortunately, the parameters $(\alpha, \beta)$ used traditionally
have become confused by frequent reversal.
}; 

\item Clipped Waxman: $p_{i,j} = \min\left(q\, e^{-sd_{i,j}},
    1\right)$, where $s \in [0,\infty), q \in (0,\infty)$;

\item Mixed Waxman-threshold: $p_{i,j} = q\, e^{-sd_{i,j}}H( r -
  d_{i,j})$, where $s \in [0,\infty), q \in (0,1], r\in [0,\infty)$;
  
 \item Threshold: $p_{i,j} = q\,H( r - d_{i,j}) $, where $q \in (0,1],
  r\in [0,\infty)$, (motivated by the {\em random plane
    network}~\cite{gilbert61:_random_plane_networ});

\item GER: $p_{i,j} = q$, where $q \in (0,1]$
  \cite{gilbert59:_random_graph,erdos60};

\item Power law: $p_{i,j} = q\, {(1 + \theta_{1} \, d_{i,j})}^{
    -\theta_{2}}$, (\eg {\em range-dependent random graphs})
  \cite{grindrod02:_range_proteom,farago02:_scalab_analy_desig_ad_hoc,gunduz-demir07:_mathem,hekmat3:_ad_networ}

\item Cauchy: $p_{i,j} = q\, {(1 + \theta_{1} \, d_{i,j}^{2})}^{ -1}$
  \cite{Avin:2008:DGR:1400863.1400878},

\item Exponential: $\displaystyle p_{i,j} = \frac{ q\, { e^
      {-d_{i,j}}}} {L - d_{i,j}}$,
  \cite{zegura97:_quant_compar_of_graph_based};

\item Max entropy: $\displaystyle p_{i,j} = \frac{
    q e^{-sd_{i,j}}} { 1 + q\,e^{-sd_{i,j}}}$;

\end{itemize}
where $H(a)$ is the Heaviside step function, and $L$ denotes the
longest possible link in the region in question. Note that our
parameterizations are sometimes different from those in the
literature so that they can be presented consistently.

All examples of which we are aware  have non-increasing link-probability
functions. We refer to these as {\em distance deterrence
  functions}, and exploit this property in our algorithm.

Many of the properties of SERNs are known. For instance, the average
node degree in the Waxman graph is \cite{roughan15:_estimwaxman}
\begin{equation}
  \label{eq:node_degree}
  \bar{k} = (n-1) q \tilde{G}(s),
\end{equation}
where $\tilde{G}(s)$ is the Laplace transform of $g(t)$, the
probability density function between an arbitrary pair of random
points (as in the Line-Picking Problem)
\cite{b.ghosh51:_random_rect,Rosenberg200499}.

\subsection{Fast generation of GER graphs}

The common method for generation of GER $G_{n,p}$ is simply to perform
$O(n^2)$ Bernoulli trials, one for each possible edge. 
Batagelj and Brandes \cite{PhysRevE.71.036113} noted that this
algorithm is \naive and that a faster implementation was possible.

The best algorithm is $\Omega(n + e)$, \ie no SERN generation
algorithm can be faster than a factor of the number of nodes and edges
generated, because the edges are independent (conditional on node
locations).

Batagelj and Brandes \cite{PhysRevE.71.036113} noted that if we list the
possible edges, then taking sequential trials generates a discretized
Poisson Process. If the graph is sparse, then it is much faster to
generate the points of this process by taking geometrically
distributed jumps. The result is an $O(n + e)$ algorithm as it
generates one edge per jump. 


Somewhat surprising, particularly as \cite{PhysRevE.71.036113}
presents fast methods for generating $G(n,m)$, we have seen no
works that consider generation by using the fact that $G_{n,p} = G(n,
M)$ where $M \sim Binomial(\ell, p)$, where $\ell$ is the number of
possible edges. In this case we might generate $M$ from the binomial
distribution (or more efficiently for sparse networks via its Poisson
approximation), and then use the resampling technique of
\cite{PhysRevE.71.036113} which is $O(m)$ to generate the
network. This evidently scales just as the previous algorithm, but the
constant time components are different, and so for some parameter
values this approach might be faster.


Regardless, our goal here is to exploit some of these ideas to
generate (sparse) SERNs, but it is not so simple: we can not
just generate a jump process, because all links are not equal, and
likewise we can not sample from the possible links.  However, 
in addition to the insight of Batagelj and Brandes we add
that the jump process on the edges allows for the edges to be listed
in any order. That means we can exploit the geometrical structure of
the SERN to list the possible edges in an advantageous order for
synthesis. 

\section{Fast Waxman Generation} 
\label{fastgen}

All SERN generators start by generating a set of $n$ nodes, which
takes $O(n)$ operations. We discuss methods for doing so quickly and
efficiently in \autoref{sec:implementation}, as this requires
some implementation tricks. Here we concentrate on the main
performance bottleneck, which is generating the edges.

For simplicity, we describe our edge generation algorithms here
specifically for the Waxman SERN, though our code generalizes this for
all the cases described above.  The {\em \naive algorithm} for
generating the edges of an undirected Waxman graph is shown in
\autoref{alg:naive}.

\begin{algorithm}
  \begin{algorithmic}[1] 
    \STATE Input: $n$, $q$, $s$
    \STATE ${\cal E} \leftarrow \phi$
    \FOR{i = 1..n}
      \FOR{j = i+1..n}
        \STATE calculate $d_{i,j}$\;
        \STATE calculate $p_{i,j} \leftarrow q \exp( -s d_{i,j})$\;
        \STATE generate $r \sim U[0,1]$\;
        \IF{$r \leq p_{i,j}$}
          \STATE ${\cal E} \leftarrow {\cal E} \cup (i,j)$
        \ENDIF
      \ENDFOR
    \ENDFOR
  
    \caption{The \naive algorithm for generating the edges of an
      undirected Waxman graph. We refer here to the uniform random
      variate on the interval $[0,1]$ as $U[0,1]$.}
    \label{alg:naive}
  \end{algorithmic}
\end{algorithm}

Our first algorithm -- $q$-jumping -- uses the observation that
\begin{equation}
  \label{eq:q-jump}
  p(d_{i,j}) = q \, e^{- s d_{i,j}} \leq q.
\end{equation}
Thus, there exists a GER $G_{n,q}$ random graph that is an ``upper
bound'' on the Waxman random graph, in the sense that each Waxman
random graph is a subgraph of a GER random graph.  We can generate the
the GER graph using the jump process described above, and then filter
to obtain the Waxman graph as shown in \autoref{alg:q-jump}.

\begin{algorithm}[t]
  \begin{algorithmic}[1] 
    \STATE Input: $n$, $q$, $s$
    \STATE ${\cal E} \leftarrow \phi$
    \STATE ${\cal E}_1 \leftarrow G_{n,q}$
    \FOR{$(i,j) \in {\cal E}_1$}
        \STATE calculate $d_{i,j}$\;
        \STATE calculate $p'_{i,j} \leftarrow \exp( -s d_{i,j})$\;
        \STATE generate $r \sim U[0,1]$\;
        \IF{$r \leq p'_{i,j}$}
          \STATE ${\cal E} \leftarrow {\cal E} \cup (i,j)$
        \ENDIF
    \ENDFOR
  
    \caption{The $q$-jumping algorithm for generating the edges of an
      undirected Waxman graph.}
    \label{alg:q-jump}
  \end{algorithmic}
\end{algorithm}

The computational cost of the $q$-jumping algorithm can be seen to be
$O(e_1)$ where $e_1$ is the number of edges in the $GER_{n,q}$
graph. We can derive this number of edges in relation to the Waxman
graph by noting \cite{roughan15:_estimwaxman} that
\begin{eqnarray*}
  \E[e_1]  & = &  n \bar{k} / 2, \\
  \E[e]    & = &  n \bar{k} \tilde{G}(s) / 2,
\end{eqnarray*}
where  $\bar{k}$ is average node degree. The algorithm is therefore
overall $O(e)$, but we want not only good order performance, but
also efficient algorithms. The efficiency of this approach depends on
the ratio of the two expectations, $\tilde{G}(s)$.

A Laplace transform of a PDF obeys certain properties: $\tilde{G}(0)=
1$, and $\tilde{G}(s)\rightarrow 0$ for large $s$, so the $q$-jumping
algorithm will be quite efficient for small $s$, but less so as $s$
grows. On the other hand, the main property of the Waxman graph is that for
larger $s$, long links are unlikely. Thus the very nature of these
graphs creates geometric structure, we can exploit in their
generation. 

\begin{figure}[thb]
  \centering
  \includegraphics[width=0.6\figurewidth]{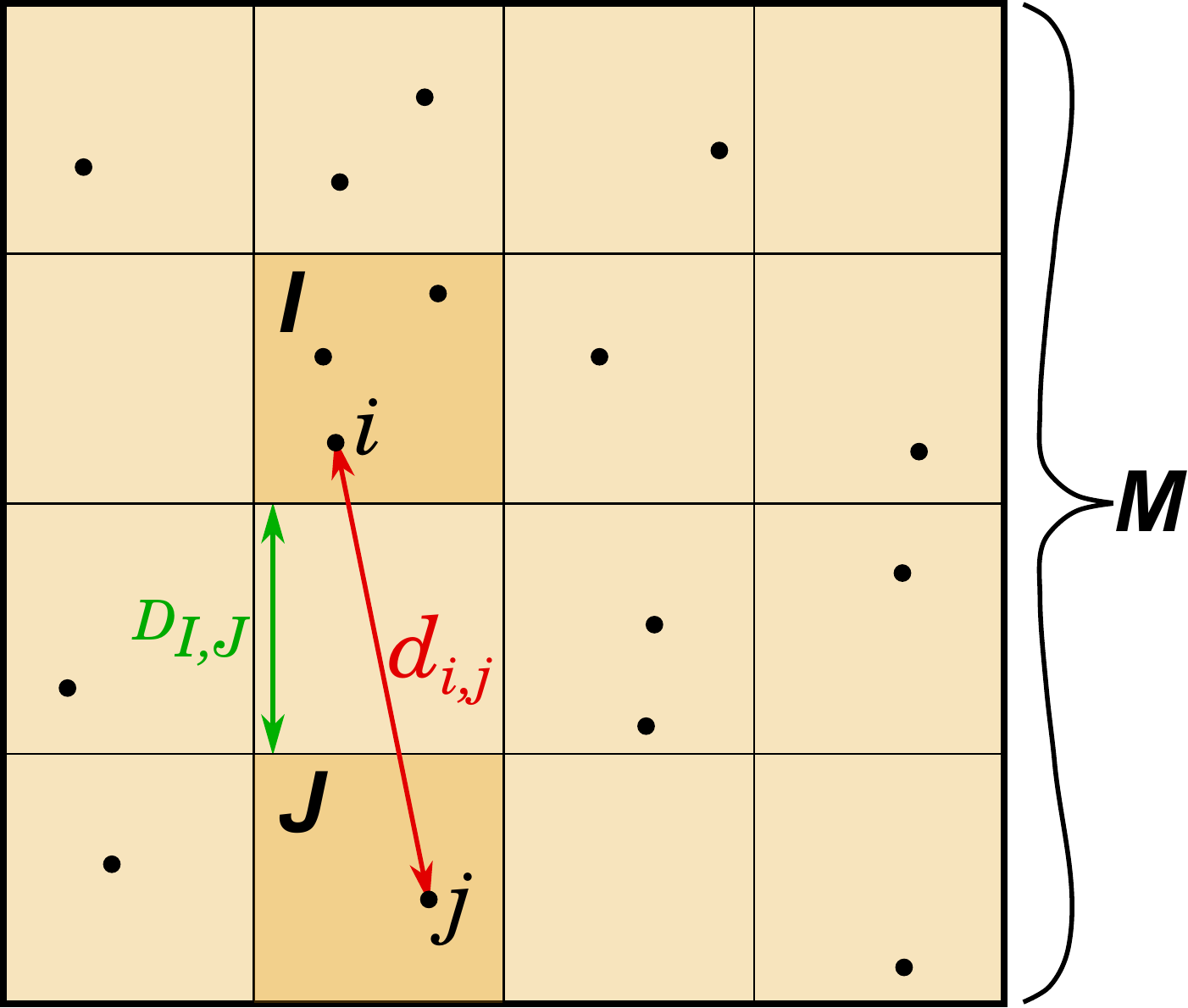}
  \caption{Region broken into buckets. We refer to these as buckets
    rather than the more obvious grid, or other terms, because in
    general they might not form a regular grid. }
  \label{fig:buckets}
\end{figure}

We do so by breaking the region into $M^2$ ``buckets'' as shown in
\autoref{fig:buckets}. Given nodes $i$ and $j$ in buckets $I$ and
$J$, respectively, we can put a lower bound $D_{I,J} \leq d_{i,j}$ on
the distance between the nodes, and thus an upper bound on the
probability of a link. 

Note that as the GER jumping algorithm does not depend on the
order of the potential edges, or even that we generated them all at
once,  we can use this approach to generate the set of edges
between any pair of buckets, using the upper bound given above. The
resulting algorithm is shown in \autoref{alg:bucket}

\begin{algorithm}[t]
  \begin{algorithmic}[1] 
    \STATE Input: $n$, $q$, $s$, $M$
    \STATE ${\cal E} \leftarrow \phi$
    \FOR{I=1..M}
      \FOR{J=I..M}
         \STATE $N_{I,J} \leftarrow $ number of possible node pairs
         \STATE $Q_{I,J} \leftarrow q \exp(-s D_{I,J})$ 
         \STATE ${\cal E}_{I,J} \leftarrow G_{N_{I,J},Q_{I,J}}$
         \FOR{$(i,j) \in {\cal E}_{I,J}$}
            \STATE calculate $d_{i,j}$\;  
            \STATE calculate $p'_{i,j} \leftarrow \exp\big( -s (d_{i,j}-D_{I,J})\big)$\;
            \STATE generate $r \sim U[0,1]$\;
            \IF{$r \leq p'_{i,j}$} 
               \STATE ${\cal E} \leftarrow {\cal E} \cup (i,j)$
            \ENDIF
         \ENDFOR
      \ENDFOR
    \ENDFOR
  
    \caption{The bucket algorithm for generating the edges of an
      undirected Waxman graph.}
    \label{alg:bucket}
  \end{algorithmic}
\end{algorithm}

The algorithm does not yet describe
\begin{enumerate}
\item creating the buckets; and 
\item data structures for efficiently storing the component edges, and
  bringing them back together at the end. 
\end{enumerate}
These are necessary to create fast code but do not affect the
asymptotic performance of the algorithm, which is again $O(e)$, but
faster than the $q$-jumping algorithm for large $s$. We describe these
below in \autoref{sec:implementation}.

\section{Results} 
\label{results}

We test the performance of the algorithms described above using a C
implementation, for which we provide stand-alone code, library
functions, and R and Matlab bindings. We use the last to provide a
mechanism to time generation through Matlab's \verb|tic()/toc()|
function, which provide a wall-clock time estimate which we can
compare against existing Matlab code.  We test timing by performing
100 generations and taking the shortest times for each on a Ubuntu
12.10 Linux box running on an Intel i7 X990 CPU with 6 cores running
at 3.47 GHz, with Matlab (R2013a), and gcc 4.7.2.  In each case we
generate a network with fixed average node degree $\bar{k}=1$, \ie a
sparse graph with $O(e) = O(n)$.

\begin{figure}[thb] 
  \centering
  \includegraphics[width=\figurewidth]{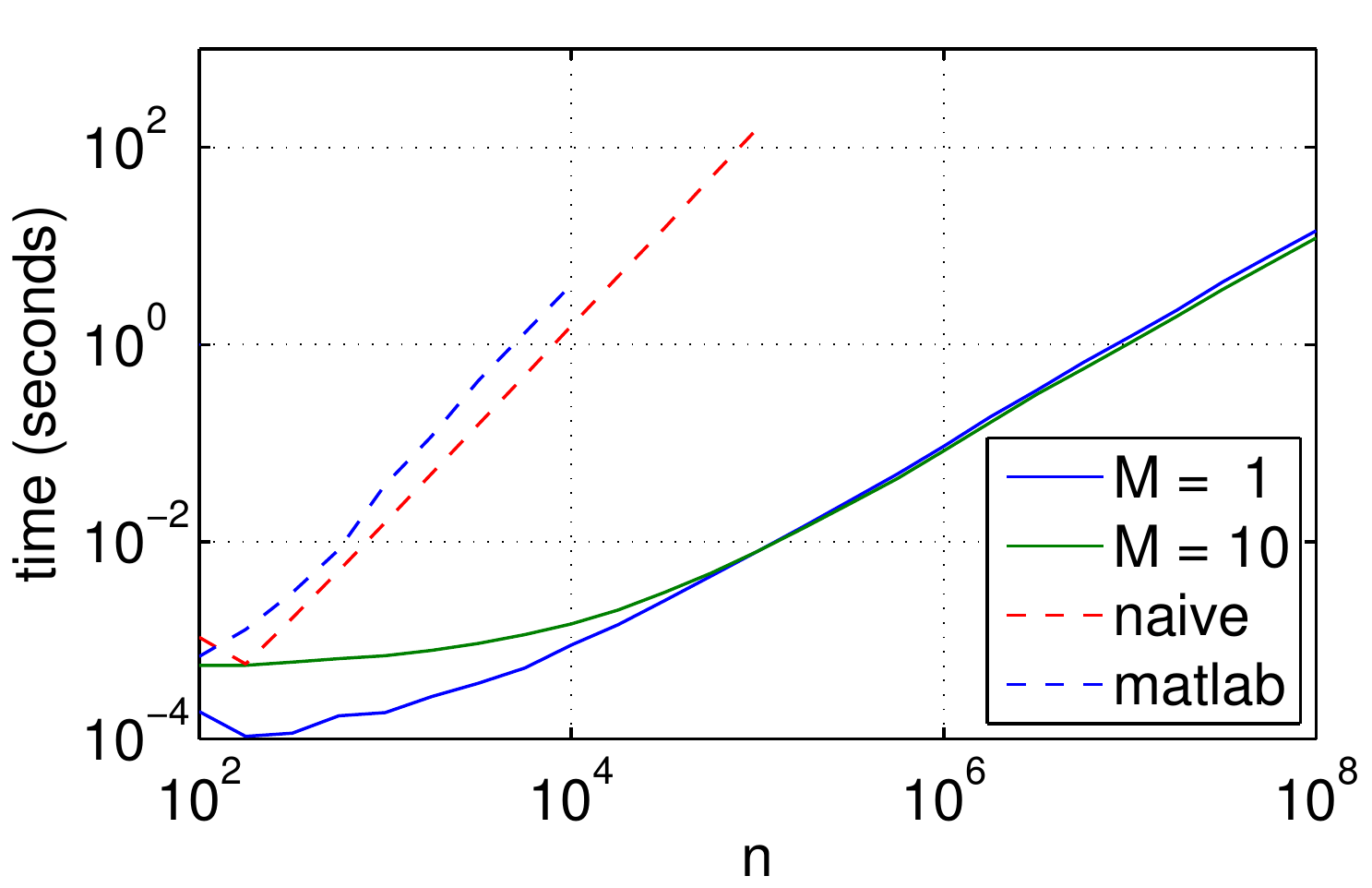}
  \caption{Performance of the bucket algorithm compared to the \naive
    algorithm for $s = 0.1$ (single thread). Note that when $M=1$ the
    bucket algorithm is equivalent to the $q$-jumping algorithm.}
  \label{fig:small_s}
\end{figure}

\autoref{fig:small_s} shows the results for a small $s$ value, over
a range of network sizes $n$, and for two bucket grid sizes $M=1$ and
10. The dashed blue curve shows results for a vectorized Matlab
implementation as a benchmark. The dashed red curve is the \naive
algorithm, which shows clear $O(n^2)$ performance, with roughly a two
times speed up in comparison to the Matlab implementation. NB: Matlab
has the capability to use multiple threads to speed up vectorized
computations, whereas this C-version uses a single thread, hence the
C-code speedup is not as great as might be expected. The Matlab
implementation uses uses $O(n^2)$ memory, so we do not try to perform
any very large tests.

The solid curves show the bucket-based algorithm for two bucket grid
sizes ($M=1$ and 10). Note that when $M=1$ the bucket algorithm is
equivalent to the $q$-jumping algorithm.

We can see for both values of $M$ that the performance for large $n$
is $O(e)$, and that the bucket grid size $M$ has negligible impact for
large $n$. For small $n$ we can see the overhead (which is $O(M^2)$)
in the initial bucket generation procedure.

\begin{figure}[thb]
  \centering
  \includegraphics[width=\figurewidth]{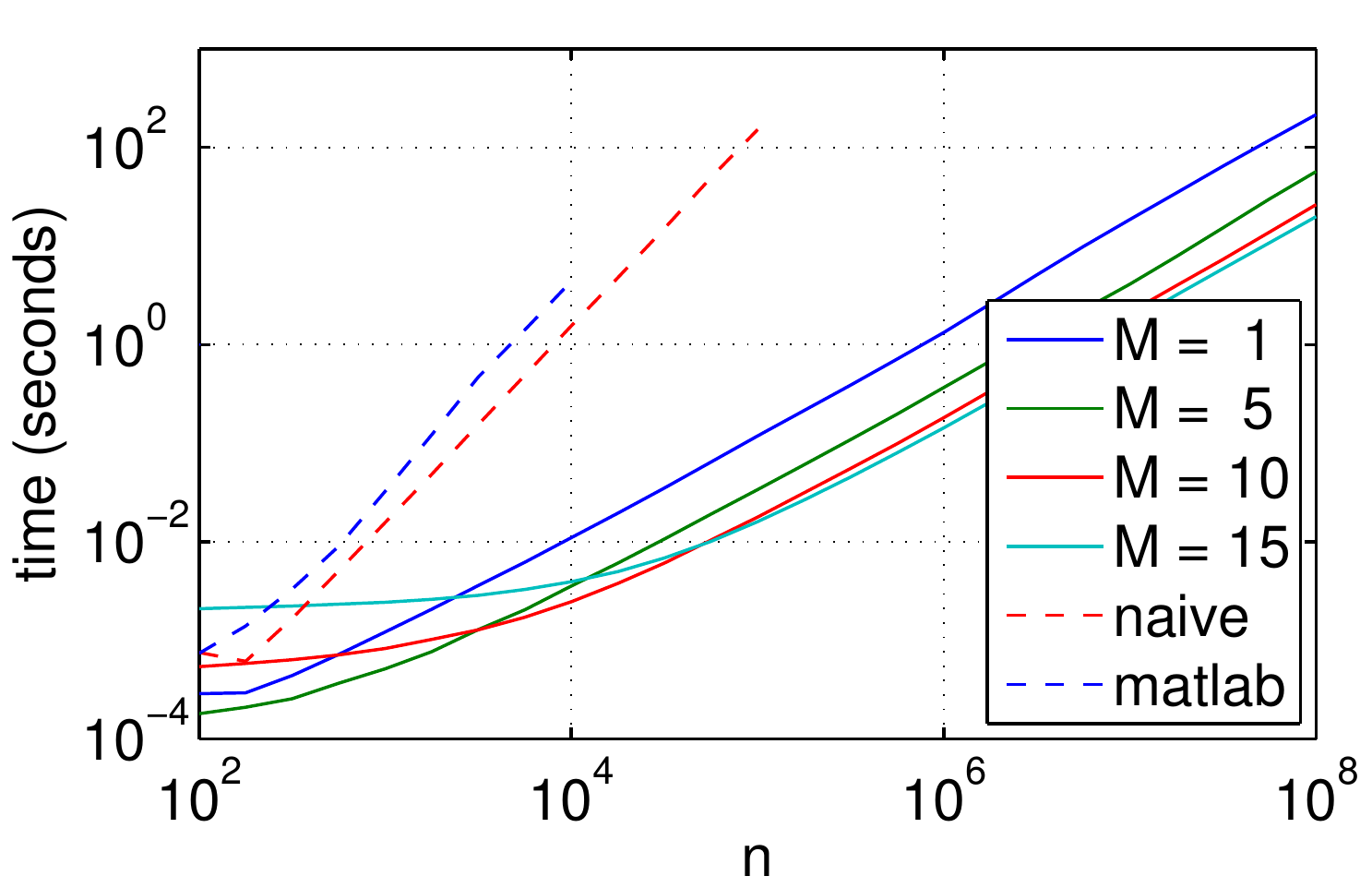}
  \caption{Performance comparison for $s = 10$ (single thread).}
  \label{fig:large_s}
\end{figure}

\autoref{fig:large_s} shows the performance for large $s$, and
although we see the same broad features as in the previous figure, we
now also see the benefit of the bucket. A larger number of buckets
improves the algorithm for larger $s$, though there is a diminishing
return as $M$ increases. 

\begin{figure}[thb]
  \centering
  \includegraphics[width=\figurewidth]{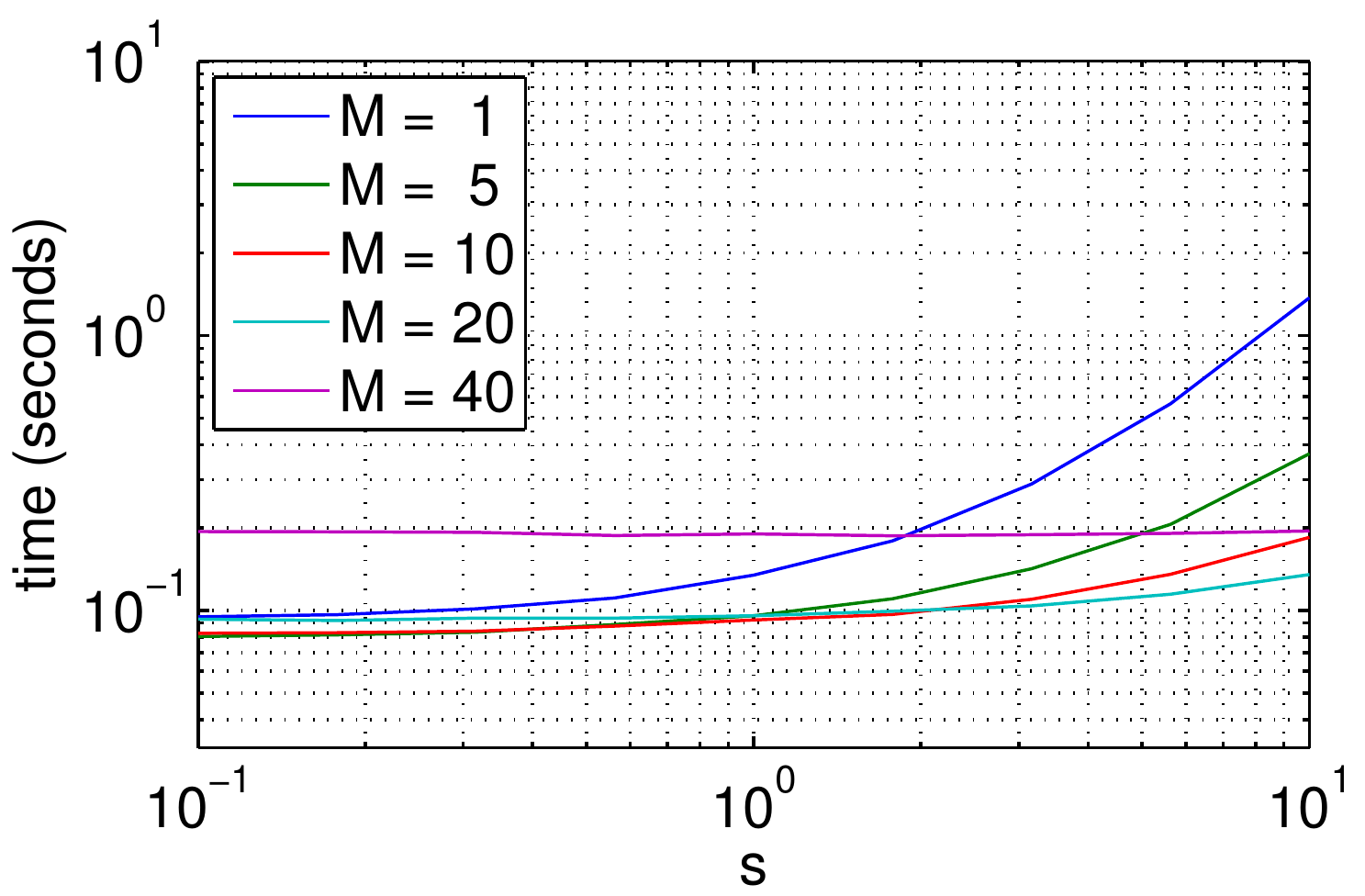}
  \caption{Performance as a function of $s$ ($n=10^6$, single thread).}
  \label{fig:M} 
\end{figure}

We consider the effect of bucket dimension more carefully in
\autoref{fig:M}, which shows the performance for fixed $n$ over a
range of $s$ values, for different bucket dimensions. Most obviously,
any fixed number of buckets has a ``sweet spot'' where it best
balances the initial overhead of bucket creation with the performance
drop-off as $s$ increases. However, a relatively small number of
buckets (around $M=20$) provides good performance over a very wide
range of parameters (note the log scales). 

\begin{figure}[thb]
  \centering
  \includegraphics[width=\figurewidth]{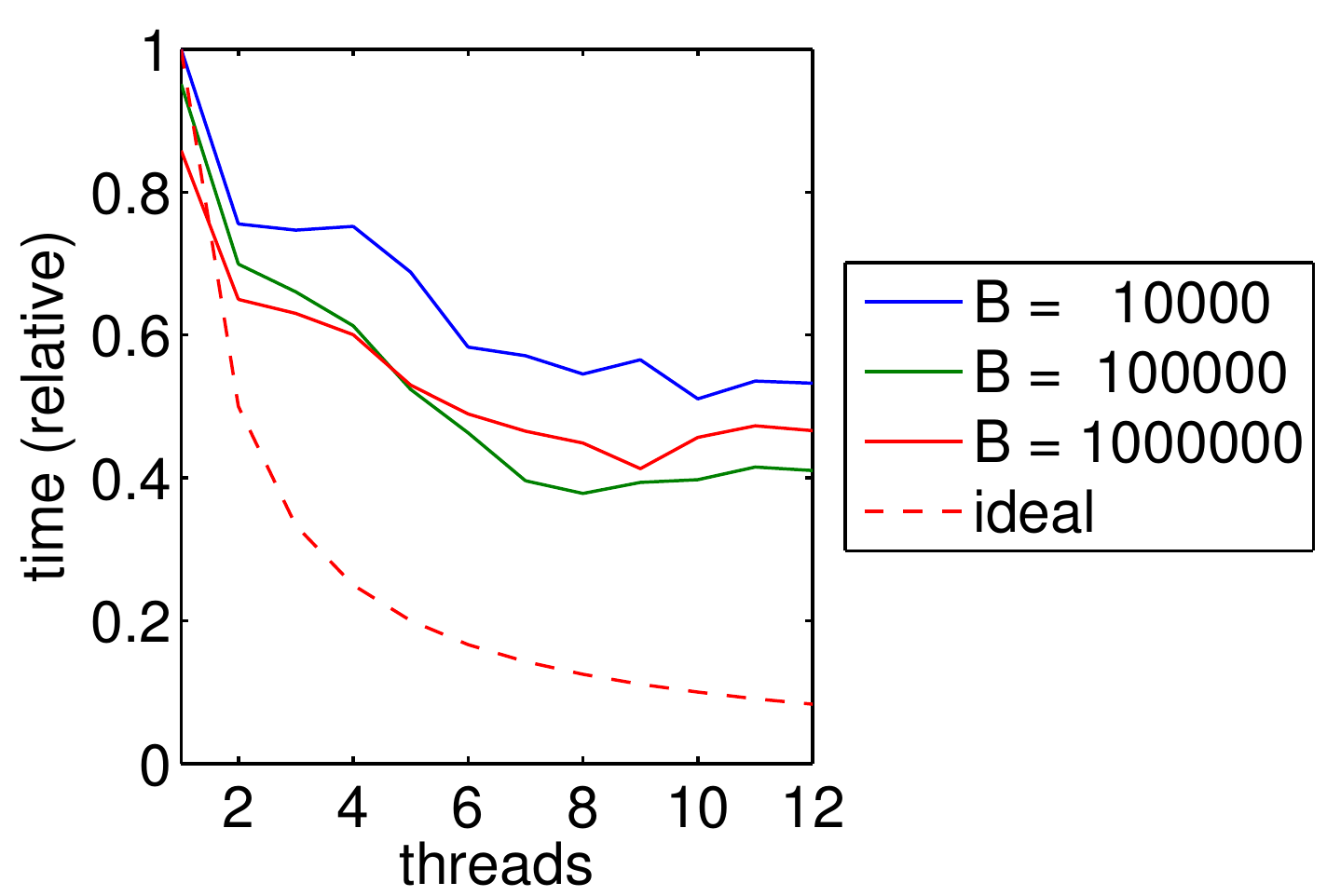}
  \caption{Multi-thread performance ($n=10^6$, $M=20$) for different
    thread-buffer sizes $B$.}  
  \label{fig:multithread}
\end{figure}

The final results shown in \autoref{fig:multithread} show the
multi-threading performance in comparison to the ideal parallelized
performance. The figure shows that the parallelization works, but that
the multi-thread implementation has significant overhead in bringing
the edges back together. If one were aiming to calculate statistical
properties of the graph that did not require it to be stored as a
whole (for instance, average node degrees or link distances), then one
could construct the information required, in parallel, without this
overhead, and thus attain the ideal performance. 
 
\section{Implementation details}
\label{sec:implementation}

Our implementation is based on a shared C library which can be built
using the supplied makefiles on Linux or Xcode project file for
OSX, using only standard C libraries already present. 

The makefiles in the package will create the following:
\begin{itemize}

\item a command line based application that outputs a GraphML file
  \cite{brandes13:_graphml}; and
  
\item a shared library, suitable to for linking to a program written
  in any language that can support the C call by reference calling
  convention.
  
\end{itemize}
The shared library is linkable with any other software capable of
using the C call by reference calling convention: we provide as
examples
\begin{itemize}
\item Matlab MEX bindings,
  
\item Rcpp based R bindings, 
  
  
\end{itemize}


High-level tools such as as Matlab and R do their own garbage
collection, so data passed back to them must be allocated using
specific functions that allow access by the callers' garbage
collection routines. Rather than conditionally compile the routines on
a per application basis they can take pointers to functions for
allocating and reallocating memory and will use these (if supplied) to
create any memory that will be returned to the caller.

The implementation was developed under OSX and Linux concurrently and
the thread model chosen was the POSIX \texttt{pthread} library as it
was well supported on both.

The aim is to create very large graphs so the use of memory is
important. We avoid the use of data structures which fragment
memory and thus cause the processor to try to cache
memory from both ends of the available address space
simultaneously. Using linear data structures maximizes the
effectiveness of the cache.

We achieve very close to lower bound on the amount of memory needed to
fully specify a SERN. Other than the lookup table $Q$ and a small
buffer for each bucket, we store only the $(x,y)$ coordinates of the
nodes and $(from, to)$ pairs of node identifiers that define the
links. We do not store labels for each node as they have an implicit
ordering.

Our code has an upper limit of $2^{32}$ on the number of nodes $N$, so
links $(i,j)$ can be represented as two 32 bit integers. Coordinates
$(x_i,y_i)$ and optionally the distances for each link $d_{i,j}$ are
represented as 32 bit floats. So the total storage requirements
are 
\begin{itemize}

\item $(x_i,y_i)$ for $i=1,\ldots,N$ takes $2 \times 4 \times n$
  bytes;

\item ${\cal E} = \{ (i,j) \}$ takes $2 \times 4 \times e$ bytes; and

\item {\em optionally} $d_{i,j}$ for $(i,j) \in {\cal E}$ takes $1
  \times 4 \times e$ bytes.

\end{itemize}
Total memory usage is $8(n+e)$ to store the graph, or $8 n + 12 e$
bytes if we include distances.  In the largest example we have
considered, with $n = 10^9$ and $E[k]=3$, the memory usage was around
$20$ gigabytes (measured in powers of ten).

The implementation has two phases:
\begin{itemize}

\item Node generation within bucket data structures given the
  restrictions of the geometry used to define the region; and

\item Link generation following algorithm described above.

\end{itemize}
The following sections give an overview of the techniques used in each
phase.

\subsection{Node generation}
\label{nodegen}

The running time of our algorithm is dominated by the time spent
creating links, but most of the code is devoted to setting the
preconditions for the algorithm to work efficiently. We need to make
the discovery of the connection buckets and the nodes they contain an
$O(1)$ operation.  Our implementation does this by creating an array
of bucket data structures, each of which record a node count and an
offset to the start of its data in the arrays of $x$ and $y$ values
associated with the nodes. Whilst this arrangement is straight forward
to produce na\"{i}vely, particularly when the SERN is embedded in a square
region, our implementation allows for parallel execution and three
types of region in which to embed the SERN:
\begin{itemize}
\item A rectangular region, generalizing the square region initial
  investigated by Waxman \cite{b.m.waxman88:_graph};

\item An elliptical region allowing investigation of SERN's where
  there is no corner effects; and

\item A user defined polygon allowing real world boundary data to be
  used.
\end{itemize}
The technique is to find a rectangular area that covers the defined
region, and then divide this into $M_{1}\times M_{2}$ square buckets,
where we denote $M = max(M_{1},M_{2})$. 

We then calculate the area $A_{I,J}$ of the intersection of the
defined region with bucket $(I,J)$.
\begin{itemize}
\item  For non-square rectangular regions we simply compare the
  dimensions of the defined region with the dimensions of the area
  covered by the $M_1 \times M_2$ square buckets,

\item For elliptical regions if all the points defining a bucket lie
  wholly inside or outside the region then $A_{I,J}$ is the area of
  the bucket or zero respectively. If the bucket intersects the
  boundary of the region then the method described by Groves
  \cite{groves1963area} is used to calculate the intersection area.

\item For a region defined using a polygon if the bucket is wholly
  inside or outside the defined region the $A_i$ are as in the case of
  an elliptical region. If the bucket intersects the boundary of the
  region then the Sutherland-Hodgman~\cite{sutherland1974reentrant}
  algorithm is used to calculate the intersection area.
\end{itemize}
At present, these precalculations are done per call, but it is
clear that if more than one graph is to be generated on the same
region, these could be precalculated once. 

The probability of a node being in a particular bucket is $P_{I,J} =
{A_{I,J}}/{A}$. We calculate the number of nodes in each bucket in
advance using a multinomial distribution $Mult(n, P_{I,J})$ to allow
memory allocation to be performed once. That makes node creation and
allocation to buckets \emph{embarrassingly parallel}. Also, all nodes
can be stored in a single contiguous memory block with a separate
pointer to the start of each bucket, rather than separate memory for
each bucket.

The algorithm for generating multinomials is simply the conditioned
repeated use of the algorithm for generating binomial random variates
due to Knuth \cite{knuth1968art}.  

The resultant output is an array of $M_1 \times M_2$ bucket data
structures containing a count of the number of nodes within them and
offsets into the larger array containing the $x$ and $y$ coordinates
of the nodes. The set of nodes are not mixed or overlapped between
buckets within this array, meaning the link creation algorithm can
operate on them blindly.

To place nodes on non-square regions we generate candidates on
a square that covers the region and then accept or reject based on
their membership of the region. The rejection rate can be controlled
via the bucket size. 

The only other prerequisite for the operation of the fast link creation algorithm is the generation of a lookup table $Q$ (Algorithm 3 step 6) to enable filtering based on the distances between the buckets. $Q$ is implemented as an array where $Q[n,m]$ represents the probability of a link of the minimum length between buckets that are $n$ buckets apart in the $x$ axis and $m$ buckets apart in the $y$ axis. The discussion of distance and link probability calculation is dealt with hereunder as it also applies to the link generation phase.

In early development it was apparent that the C stdlib  library  random number generating functions have a number of limitations in this setting. The  \texttt{srand} function is not re-enterant nor is it thread safe because it stores state internally thus can be immediately disqualified. The  \texttt{rand\_r} function and  \texttt{drand48\_r} functions allow the caller to provide storage space for the state but here we will be calling the function  $2N$ times to create the coordinates for $N$ nodes and the overhead of a function call becomes substantial.  Based on the work of Marsaglia \cite{marsaglia2003random} we implemented a multiply with carry (MWC) random number generator with separate state for each thread and because it is not a library function we were able to use the C99 inline calling convention to eliminate the overhead of a function call. This random number generator is the source for creating the random variates drawn from geometric and binomial distributions used in all sections of the implementation. 

\subsection{Link generation}
The general algorithm for fast link generation has been covered in
\autoref{fastgen} so here we will use the notation of
\autoref{alg:bucket} to discuss implementation details.

The implementation combines steps 7 and 8 of the algorithm by looping
from 0 to $N_{I,J}-1$ in skips of length $X \sim Geo(q\, Q_{I,J}
)$. The sum of these skips $k$ represents a pair of nodes in $N_{I,J}$
that forms an element in ${\cal E}_{I,J}$.

It is important to be able to decode quickly which nodes in our bucket
data structure $k$ represents.  We have a different way of decoding
$k$ if $I=J$ or $I \neq J$, so the obvious implementation would
include an {\tt if} statement within the loop to determine which
decoding scheme to use. However this results in very poor execution
times compared to separating into two loops, one that only deals with
nodes from the same bucket and one that deals with nodes from distinct
pairs of buckets. This is because modern processors have long pipe
lines capable of executing more than one instruction per clock cycle,
but any branching instruction causes the pipeline to be flushed. With
such a tight loop this means that a lot of the loop is flushed out of
the pipeline on every iteration.

When the buckets $I$ and $J$ are distinct then $|N_{I,J}| = |{\cal
  N}_{I}| \times |{\cal N}_{J}|$ and decoding $k$ is straightforward
with $i = k \mod |{\cal N}_{I}|$ and $j =\lfloor{k / |{\cal
    N}_{I}|}\rfloor$.

When the nodes involved come from the same bucket then $|N_{I,J}| =
|{\cal N}_{I}| \times ({\cal N}_{I} - 1) / 2$ and $k$ is decoded using
$j = 1 + ((\lfloor \sqrt{8 k + 1}\rfloor - 1) / 2)$ and $i = k - j
(j - 1) / 2$.

The random number generator described in the \autoref{nodegen} is
used along with time-costly inverse transform to give the
geometrically distributed random variables. We move part of this
transform outside the loop as it does not rely on the random number
generated giving another performance increase.

The parallel execution of link generation is more complex than that of
node generation because we do not know in advance how many links will
be generated and we want to maximize the size of the network that we
can create in a given amount of memory. We also want to be able to
return the data in a contiguous block of memory so that applications
like Matlab and R can use it trivially. Schemes of node generation
where each thread writes to its own memory until all links are
generated are thus not possible, as they require either much more
memory or a juggling act where some sets of data are shrunk and others
grown until all data is in a contiguous space. The former is
restrictive and the latter likely to end up with a fragmented heap and
deadlock because it is not possible to transition to the next
state. The implementation chosen is a compromise with each thread
generating the links for a pair of buckets at a time and writing this
data into a buffer which when full is written in to some shared memory
using a common pointer to the next available free location. If the
number of links were known in advance this could be implemented in a
lock free manner using an atomic compare and swap (CAS) instruction,
but there is always the possibility that the memory for the links
might need to be grown forcing the use of a lock on the section where
we test if any growth is needed. The possibility of another thread
still writing its buffer whilst the memory is being grown is handled
using a CAS instruction. It is this area of the implementation that
would currently respond best to optimization with a number of
alternatives under consideration.

\section{SERNS in general}
\label{sec:general}

As noted, the current code allows generation on rectangles, ellipses
and arbitrary polygons on $\R^2$.

Currently our standard wrappers implement the four metrics and nine
link probability functions described in \autoref{sec:formalities},
but the routines for both bucket and link generation have been
parameterized to accept pointers to functions that implement distance
and link probabilities, so creating new models involves only five or
six lines of code.
 
Taking all the combinations of space, metric and probability functions
we get a total of nearly 100 different models implemented in the
default code, with the ability to add new models simply by passing
function handles, so our code can generate a very large class of
SERNs.

The major limitation at present is that the points lie in $\R^2$, but
the bucket algorithm extends to $\R^n$ and to surfaces other than the
plane, such as the sphere and cylinder. It is our intention to add
these as possibilities, but this requires additional complexity in the
generation of the buckets and management of memory. Moreover, at
present we have no clear application for SERNs in high dimensions.

\section{Conclusion and Future Work}

This paper describes an algorithm to perform fast $O(n+ e)$
generation of SERNs. The results from the implementation described show that 
the performance is several orders of magnitude faster than competing code for large graphs. 

However, there are still many improvements that could be made, and are
the topic of current work:
\begin{itemize}
\item Better threading by predicting the amount of links that will be
  generated and allocating memory in advance with sufficient leeway to
  allow for the variance in the number of links so that we can pick an
  acceptable failure rate. The algorithm could then run at whatever
  bandwidth the memory is capable of for the vast majority of the time
  and would only have to restart occasionally due to insufficient
  memory being allocated.

\item The current implementation of polygon shapes can be optimized by
  recursively finding the intersection of buckets with the defined
  region. Rather than compare the whole region with each bucket we can
  compare an arbitrary small portion of the region with each bucket.

\item Portions of the code are suitable for moving to a graphics
  processing unit (GPU) which would speed execution of these sections
  but reduce portability.

\item The current implementation is multi-threaded, but the algorithm
  is suitable for a multi-processor implementation, for instance using
  map-reduce. 

\item We aim to extend the approach to higher-dimensional spaces, and
  non-Euclidean manifolds such as the surface of a sphere.

\end{itemize}


\section*{Acknowledgements}

This work was partially supported through ARC grants CE140100049 and
DP110103505. 

\setlength{\parskip}{2mm}
\bibliographystyle{ieeetr}

\bibliography{arXiv}
 
\end{document}